\documentclass[pre,aps,showpacs,superscriptaddress,preprint]{revtex4}
\usepackage{graphicx,color}
\definecolor{red}{rgb}{1.00,0.00,0.00}
\definecolor{gray}{gray}{0.60}

\begin{document}

\title{A nonlinear classical model for the decay widths of Isoscalar Giant Monopole Resonances}

\author{P.K. Papachristou}
\affiliation{Department of Physics, University of Athens, GR--15771,
Athens,
Greece}

\author{E. Mavrommatis}
\affiliation{Department of Physics, University of Athens, GR--15771,
Athens,
Greece}

\author{V. Constantoudis}
\affiliation{Institute of Microelectronics (IMEL), NCSR "Demokritos",
P.O. Box 60228,
Aghia Paraskevi, Attiki, Greece 15310 and Physics Department, National
Technical University,
Athens, Greece}

\author{F.K. Diakonos}
\affiliation{Department of Physics, University of Athens, GR--15771,
Athens,
Greece}

\author{J. Wambach}
\affiliation{Institut f\"ur Kernphysik, Technische Universit\"at Darmstadt, Schlossgartenstr. 9, D-64289 Darmstadt, Germany}

\date{\today}

\begin{abstract}

The decay of the Isoscalar Giant Monopole Resonance (ISGMR) in nuclei is studied by means of a nonlinear classical model consisting of several noninteracting nucleons (particles) moving in a potential well with an oscillating nuclear surface (wall). The motion of the nuclear surface is described by means of a collective variable which appears explicitly in the Hamiltonian as an additional degree of freedom. The total energy of the system is therefore conserved. Although the particles do not directly interact with each other, their motions are indirectly coupled by means of their interaction with the moving nuclear surface. We consider as free parameters in this model the degree of collectivity and the fraction of nucleons that participate to the decay of the collective excitation. Specifically, we have calculated the decay width of the ISGMR in the spherical nuclei $^{208}\rm{Pb}$, $^{144}\rm{Sm}$, $^{116}\rm{Sn}$ and $^{90}\rm{Zr}$. Despite its simplicity and its purely classical nature, the model reproduces the trend of the experimental data which show that with increasing mass number the decay width decreases. Moreover the experimental results (with the exception of $^{90}\rm{Zr}$) can be well fitted using appropriate values for the free parameters mentioned above. It is also found that these values allow for a good description of the experimentally measured $^{112}\rm{Sn}$ and $^{124}\rm{Sn}$ decay widths. In addition, we give a prediction for the decay width of the exotic isotope $^{132}Sn$ for which there is experimental interest. The agreement of our results with the corresponding experimental
data for medium-heavy nuclei is dictated by the underlying classical mechanics i.e. the behaviour of the maximum Lyapunov exponent as a function of the system size.
\end{abstract}

\maketitle

\section{Introduction}

The study of the decay of nuclear giant resonances (GRs) is a subject of extensive theoretical and experimental investigation \cite{BER83,HAR04,THO07}. GRs having a large degree of collectivity are offered for the study of dissipation of collective motion by single particle motion in a many body quantum system. Moreover, it is known that regular and chaotic dynamics usually coexist in nuclear excitations and chaotic dynamics is expected to dominate in GRs \cite{BOH88,ZEL96}. The latter is due to the time-dependence induced in the nucleons dynamics and to the correlations among the nucleons originating from the residual interaction. By studying the decay of GRs we can get information about the role that chaotic dynamics plays in dissipation of the collective motion. In the damping of the collective motion in GRs several mechanisms are at work at the mean field level (one-body dissipation), where the decay is into single particle motion (p-h excitation) including particle emission (escape width) and beyond (n-body dissipation - spreading width), where the decay occurs through coupling to progressively more complicated states starting from 2p-2h excitations (collisional damping). The relative contribution of these damping processes is under investigation. In this work we consider explicitly one-body dissipation and implicitly part of the n-body dissipation and investigate the influence of the underlying classical dynamics on it.

The effects of the chaotic dynamics on the dissipation of nuclear collective motion and the consequent excitation of the nucleons has been considered in a number of previous works. We mention for one body dissipation the application of the Fermi accelerator model \cite{FER49,ROM01}, the wall formula \cite{BLO78} and its generalizations and computer simulations of classical and quantal systems of noninteracting nucleons colliding with a moving boundary undergoing periodic adiabatic oscillations \cite{BLO95}. In these earlier works the coupling of the slowly moving collective degree of freedom to the fast moving independent particle degrees of freedom has been considered disregarding self-consistency and energy conservation. Subsequently, the damping of the collective motion coupled self-consistently, ensuring energy conservation, to the single-particle motion has been studied using the Vlasov equation \cite{BAU94}. To clarify the relationship between chaos at the microscopical level and damping of collective motion, a classical version of the vibrating potential model \cite{RIN80} for finite nuclei has been considered and the case of monopole oscillations along with others has been investigated \cite{BUR95,BAL98}. Several other calculations of the damping have been carried out later by exploring the phase space \cite{DRO95}, by considering two-body interactions \cite{PAP00,RAD06} etc. 

We focus on the decay of the Isoscalar Giant Monopole Resonances (ISGMR) (breathing mode, where neutrons and protons oscillate in phase) which are also useful for the extraction of compression moduli of nuclei $K_A$ and from them the compressibility of nuclear matter $K_\infty$ \cite{BLA80}. Recent and ongoing experimental activity at different facilities including the cyclotron at Texas A\&M and at RCNP at Osaka supplement and extend the previous data clarifying the relation between the symmetry energy and $K_A$. We will consider initially the data for the ISGMR of the spherical medium-heavy nuclei: $^{208}\rm{Pb}$, $^{144}\rm{Sm}$, $^{116}\rm{Sn}$ and $^{90}\rm{Zr}$ \cite{YOU04a,YOU04,ITO03} and those for the $^{112}\rm{Sn}$ and $^{124}\rm{Sn}$ isotopes \cite{LUI05}. There has been a large number of theoretical approaches for the study of the energy and the decay width of ISGMR \cite{SPE91,COL04,PAA07}. From these studies we mention self-consistent Hartree-Fock plus Random Phase Approximation (RPA) with Gogny \cite{BLA95,PER05} or Skyrme \cite{COL92,COL04,HAM97,AGR04} or unitary correlation operators \cite{PEG06} effective interactions and self-consistent relativistic mean field plus RPA calculations \cite{VRE99,PIK01,VRE05} as well as calculations that go beyond mean field approximation, i.e. Extended RPA \cite{PEG07}, relativistic RPA with particle-phonon coupling \cite{LIT07}. In the latter reference, by extending the covariant density functional theory with the consideration of quasiparticles dressed with a cloud of low-lying collective excitations, the influence of the phonon coupling terms to the damping of GRs has been studied and led to noticeable fragmentation and spreading width. The decay widths extracted from all the above calculations are smaller (although to a different extent) from the experimental ones.

In this work we investigate the role that chaoticity plays in the decay of ISGMR of the above mentioned spherical nuclei focusing mainly on one-body processes and using a classical model that exhibits chaotic behaviour and that is based on the model of Refs. \cite{BUR95,BAL98}. Chaoticity in relation with the decay of the ISGMR has also been considered in Refs. \cite{VRE99,LAL03} in the context of the relativistic mean-field model.

The considered model is a simple classical Hamiltonian system which consists of particles moving in a Woods-Saxon well with an oscillating surface. This oscillation represents the ISGMR state in which the nuclear surface is oscillating radially, i.e. the spherical symmetry of the nucleus is preserved. The particles can interact with the surface and can exchange energy with it. The collective variable which describes the motion of the nuclear surface appears explicitly in the Hamiltonian of the system as an additional degree of freedom, therefore the total energy of the system is conserved. Moreover, in our study we take into account the escape of nucleons from the oscillating well. Our model therefore takes explicitly into account one-body dissipation (Landau damping and particle decay) and partly many-body dissipation. Chaoticity is introduced by the time-dependence of the potential and by the indirect coupling of the motion of the nucleons by means of their interaction with the oscillating nuclear surface, which gives rise to some of the correlations among nucleons that are due to the residual interaction. In Refs. \cite{PAP05,PAPTH} our model has been used to elucidate the relationship of relaxation dynamics with the structure of the phase space of the system. We calculate the decay width of the ISGMR in the spherical nuclei $^{208}\rm{Pb}$, $^{144}\rm{Sm}$, $^{116}\rm{Sn}$ and $^{90}\rm{Zr}$. It is found that under the appropriate assumptions the trend of the experimental data can be reproduced, i.e. the decay width is a decreasing function of the mass number A. Moreover, with the proper choice of the free parameters, a good quantitative agreement with the experimental results can be obtained. This is attributed to the dependence of the mean value of the maximum Lyapunov exponent on the size of the nucleus. With the same choice of the parameters the decay width of three other isotopes of $\rm{Sn}$ ($^{112}\rm{Sn}$, $^{124}\rm{Sn}$ and $^{132}\rm{Sn}$) is calculated. Finally, the fraction of the width due to the escape is deduced for the above nuclei. In Sec. II we present a brief description of the model and the method of calculation. In Sec. III our results are presented and discussed whereas in Sec. IV the conclusions are drawn amd some prospects for further study within our model are mentioned.

\section{Description of the model and method of calculation}
The model we consider consists of a system of $N$ noninteracting nucleons (particles) of mass $m$ moving in a Woods-Saxon well. The nucleons exchange energy with the nuclear surface (wall of the potential well), which is considered to move in a harmonic oscillator potential with angular frequency $\omega$ \cite{BAL98,BUR95,PAP05,PAPTH}. A mass $M$ is assigned to the nuclear surface and its motion around the equilibrium position $R_0$ is described by the collective one-dimensional degree of freedom $R$. Since the potential well is of finite depth, nucleons can escape from it yielding a contribution to the decay width. The Hamiltonian of the system equals
\begin{equation}
H = \sum\limits_{i = 1}^N {\left[ {\frac{{p_{ri}^2 }}{{2m}} - \frac{{V_0 }}{{1 + \exp \left( {\frac{{r_i  - R}}{a}} \right)}} + \frac{{L_i^2 }}{{2mr_i^2 }}} \right]}  + \frac{{p_R^2 }}{{2M}} + \frac{1}{2}M\omega ^2 \left( {R - R_0 } \right)^2 + \frac{L_R^2}{2mR^2}.
\end{equation} 
The angular momentum $L_R$ of the collective degree of freedom $R$ is taken equal to zero, i.e. the nuclear surface is not rotating. The angular momenta $L_i$ of the particles are constants of the motion since the oscillating potential retains its spherical symmetry. The values of the parameters we use equal $a=0.67 fm$, $V_0=45 MeV$ and $R_0=1.2A^{1/3}fm$. The above values of $a$ and $V_0$ correspond to the nucleus $^{208}Pb$. There is a slight variation of $a$ and $V_0$ with the mass number $A$ but for the present study it is neglected. For the angular frequency $\omega$ we use the frequency of the ISGMR obtained from the experimentally measured centroid energy $E_x$. The values of the frequency we use for the nuclei $^{208}\rm{Pb}$, $^{144}\rm{Sm}$, $^{116}\rm{Sn}$ and $^{90}\rm{Zr}$ equal $13.96 MeV\cdot\hbar^{-1}$, $15.40 MeV\cdot\hbar^{-1}$, $15.85 MeV\cdot\hbar^{-1}$ and $17.81 MeV\cdot\hbar^{-1}$ respectively (see Table \ref{tab_gamma},\cite{YOU04a,YOU04}). The parameters $M/m$ and $N$ will be considered as free parameters and their optimal values will be determined by comparing our results to the experimental data. More precisely, free parameters are the quantities $M/mA \cdot  100\%$ and $N/A \cdot 100\%$ (see Eqs.\ref{eq_lam} and \ref{eq_xi}). The equations of motion are
\begin{equation}
\dot r_i  = \frac{{p_{ri} }}{m}\rm{~~~~~~~~}(i=1,\ldots,N)
\end{equation}
\begin{equation}
\dot p_{ri}  =  - \frac{{V_0 }}{a}\frac{{\exp \left( {\frac{{r_i  - R}}{a}} \right)}}{{\left[ {1 + \exp \left( {\frac{{r_i  - R}}{a}} \right)} \right]^2 }} + \frac{{L_i^2 }}{{mr_i^3 }}\rm{~~~~~~~~}(i=1,\ldots,N)
\end{equation}
\begin{equation}
\dot R = \frac{{p_R }}{M}
\end{equation}
\begin{equation}
\dot p_R  = \frac{{V_0 }}{a}\sum\limits_{i = 1}^N {\frac{{\exp \left( {\frac{{r_i  - R}}{a}} \right)}}{{\left[ {1 + \exp \left( {\frac{{r_i  - R}}{a}} \right)} \right]^2 }}}  - M\omega ^2 \left( {R - R_0 } \right) ,
\end{equation}
and have been solved using a leapfrog algorithm \cite{STR04}. With a simple rescaling of the equations of motion we find that the relevant parameters are $N$, $M/m$, $R_0/a$ and $V_0/(ma^2\omega^2)$.

For each nucleus we consider a microcanonical ensemble of 5000 initial conditions selected according to the following prescription:
\begin{enumerate}
\item The initial energy $E_R$ associated with the collective variable $R$ 
\begin{equation}
E_R  = \frac{{p_R^2 }}{{2M}} + \frac{1}{2}M\omega ^2 \left( {R - R_0 } \right)^2
\end{equation}
is taken equal to the experimentally measured centroid energy $E_x$ of the ISGMR. The corresponding initial momentum $p_R$  is taken equal to zero, i.e. the harmonic oscillator describing the collective motion is initially at one of its extremal points:
\begin{equation}
R = R_0  + \sqrt {\frac{{2E_x }}{{M\omega ^2 }}}.
\end{equation}
\item The initial position $r_i$ of the nucleon is uniformly distributed in the interval $[0,R]$. The nucleon is a proton with a probability $P=Z/A$ and a neutron with a probability $1-P$.
\item The initial relative energy $\Delta E_i$ of the nucleon with respect to the bottom of the potential well is selected using the density of energy of a Fermi gas, namely
\begin{equation}
n(E_i) = \frac{3}{{2E_F^{{\textstyle{3 \over 2}}} }}\Delta E_i^{{\textstyle{1 \over 2}}},
\end{equation}
where $E_F$ is the Fermi energy, which for protons is given by
\begin{equation}
E_{F(p)}  = \frac{{\hbar ^2 }}{{2m}}\left[ {3\pi ^2 \frac{Z}{V}} \right]^{{\textstyle{2 \over 3}}}
\end{equation}
and for neutrons by
\begin{equation}
E_{F(n)}  = \frac{{\hbar ^2 }}{{2m}}\left[ {3\pi ^2 \frac{A-Z}{V}} \right]^{{\textstyle{2 \over 3}}}.
\end{equation}
The volume $V$ of the system is taken equal to that of a sphere of radius $R_0$. The initial kinetic energy $T_i$ of the nucleon is therefore given by
\begin{equation}
T_i =  - V_0  + \Delta E_i + \frac{{V_0 }}{{1 + \exp \left[ {\frac{{r_i - R}}{a}} \right]}}.
\end{equation}
and its partition in radial and angular part is determined by the angle $\theta_i$ which is uniformly distributed in the interval $[0,\pi]$ ($\theta_i$ is the angle between the vectors of the initial position and initial velocity), obtaining the radial component of the momentum $p_{ri}$ and the angular momentum $L_i$ from the equations $p_{ri}  = \sqrt {2mT_i} \cos \theta_i$ and $L_i  = \sqrt {2mT_i} r_i\sin \theta_i$ respectively. We should note that we have also tried other initial momentum distributions (i.e. uniform or Boltzmann) and it has been found that the results are not sensitive to the particular choice of the initial momentum distribution.
\end{enumerate}
In our study we neglect the Coulomb barrier. However, we have carried out calculations with the Coulomb barrier for the nucleus $^{208}\rm{Pb}$ and our results for the decay width did not change appreciably. For each nucleus, we consider as free parameter the quantity
\begin{equation}
\Lambda  = \frac{M}{{mA}} \cdot 100\% ,
\label{eq_lam}
\end{equation}
which is the percentage of the mass assigned to the collective degree of freedom and consequently can be thought of as a measure of the degree of collectivity of the motion. Apart from $\Lambda$, we also consider as a free parameter the quantity
\begin{equation}
\Xi  = \frac{N}{A} \cdot 100\%,
\label{eq_xi}
\end{equation}
which measures the percentage of nucleons that participate into the damping of the collective motion. For each nucleus and for a grid of values of $\Xi$ and $\Lambda$, we calculate the mean value $<R(t)>$ of the collective variable $R$ over the selected ensemble of $N_{tot}=5000$ orbits. We also calculate the mean value of the maximum Lyapunov exponent $<\lambda_1>$, which is the most commonly used measure of the intensity of chaotic dynamics, using the same initial conditions and integration time as the ones we used in the calculation of $<R(t)>$. $<\lambda_1>$ quantifies how fast initially closed trajectories deviate in phase space and it is related to the decay width $\Gamma$ (or the relaxation time). The comparison of $<\lambda_1>$ and $\Gamma$ reveals the microscopic origins of relaxation dynamics as well as the relative importance of the regular regions in phase space \cite{PAP05,PAPTH}. In order to estimate the decay width, we take the Fourier transform of $<R(t)>$ and calculate the standard deviation $\sigma$ of the corresponding power spectrum in a frequency region around $\omega$ where the power spectrum has a maximum. In order to obtain the width $\Gamma$, which corresponds to the full width at half maximum (FWHM) of an equivalent Gaussian fit to the power spectrum, we multiply $\sigma$ by $2\sqrt{\ln 4}\simeq 2.355$. 

It seems that the use of a classical model is quite satisfactory and the violations of Fermi statistics are not so important. In particular, the effect of the spin on the momentum and energy distribution is not expected to be significant. Results are not sensitive to the type of the initial momentum distribution. In fact, the small influence of the Fermi statistics can be drawn by the small deviation of our results from the experimental data (see section III). Blocki and coworkers \cite{BLO95} have come to the same conclusion by performing computer simulations and have shown that results were similar for the excilation of a classical or quantal gas of particles by a time-dependent potential well. The model used in the present work differs from the one of Baldo et al \cite{BUR95,BAL98} mainly in the consideration of diabatic motion for the nuclear surface and realistic geometry for the nuclear potential. Moreover, in the present model the number of particles that collide with the wall as well as the mass attributed to the oscillating surface have been treated as free parameters.

\section{Results and discussion}
\begin{figure}
\begin{center}
\includegraphics[width=7.5cm]{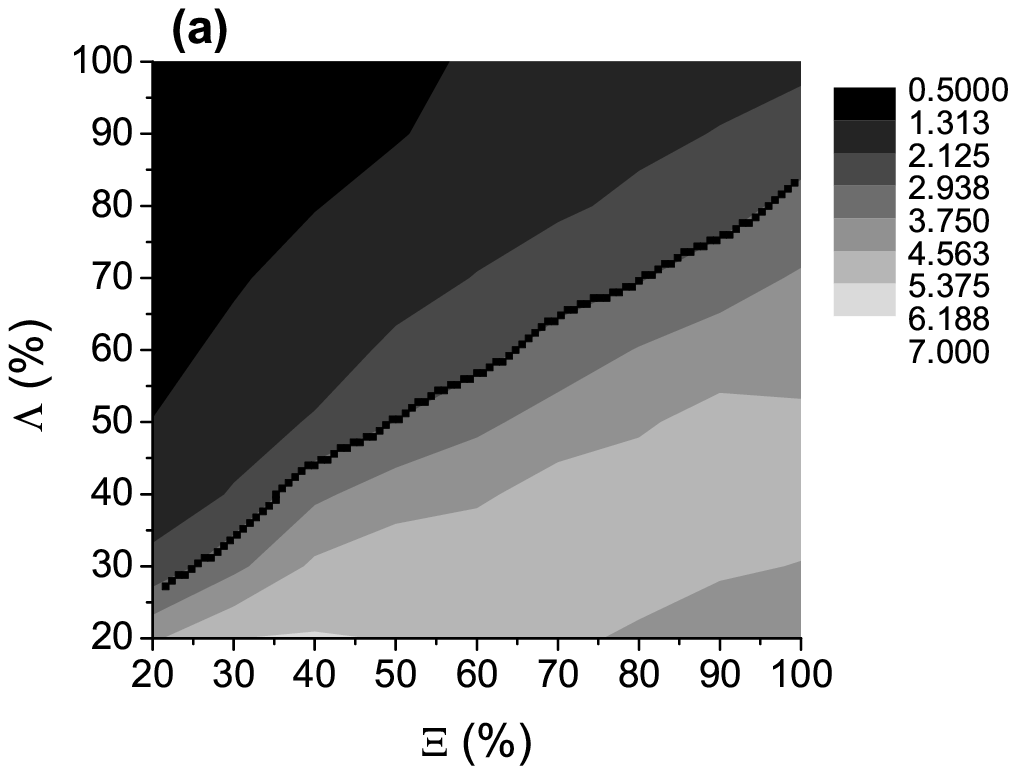}%
\includegraphics[width=7.5cm]{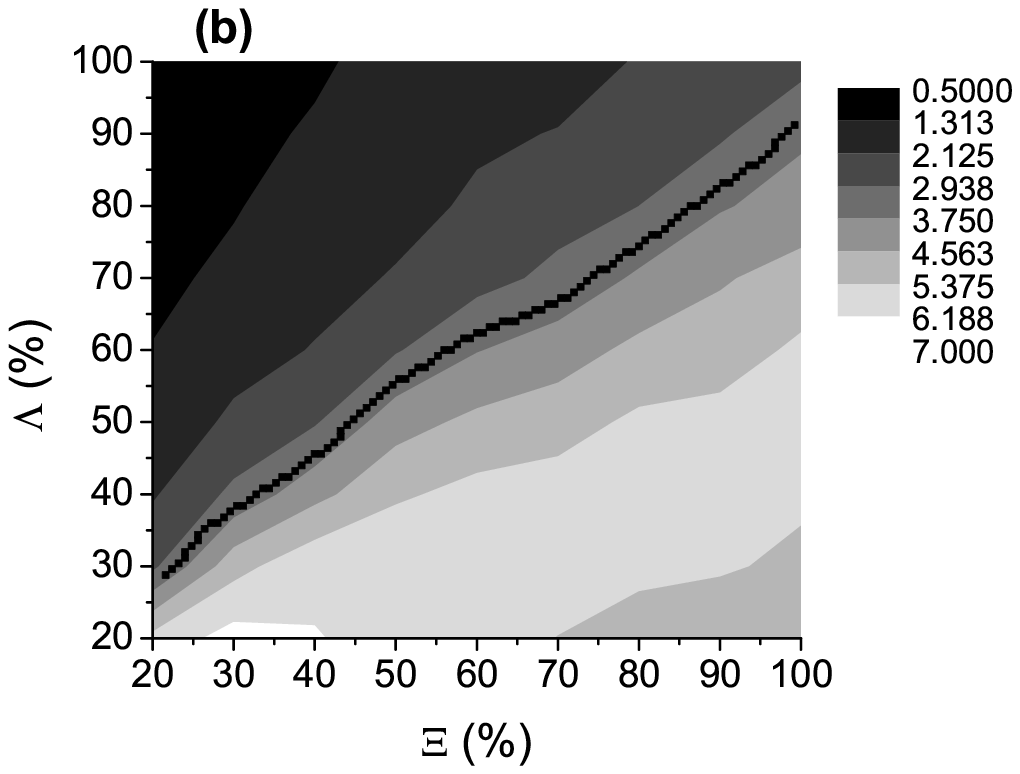}
\includegraphics[width=7.5cm]{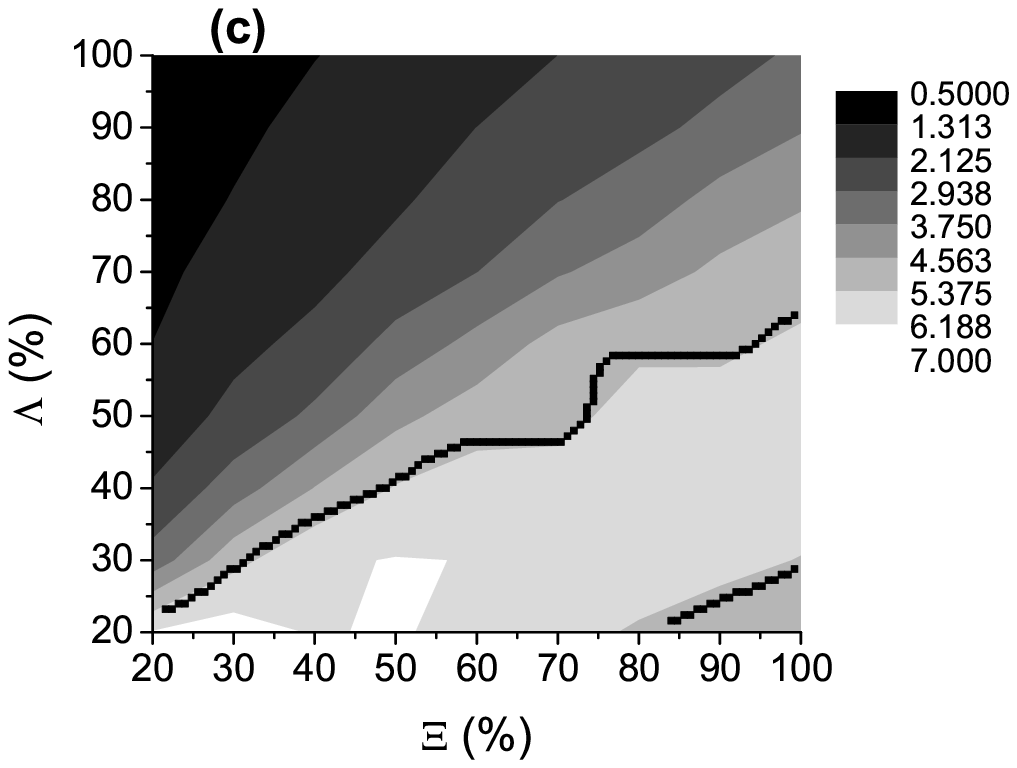}%
\includegraphics[width=7.5cm]{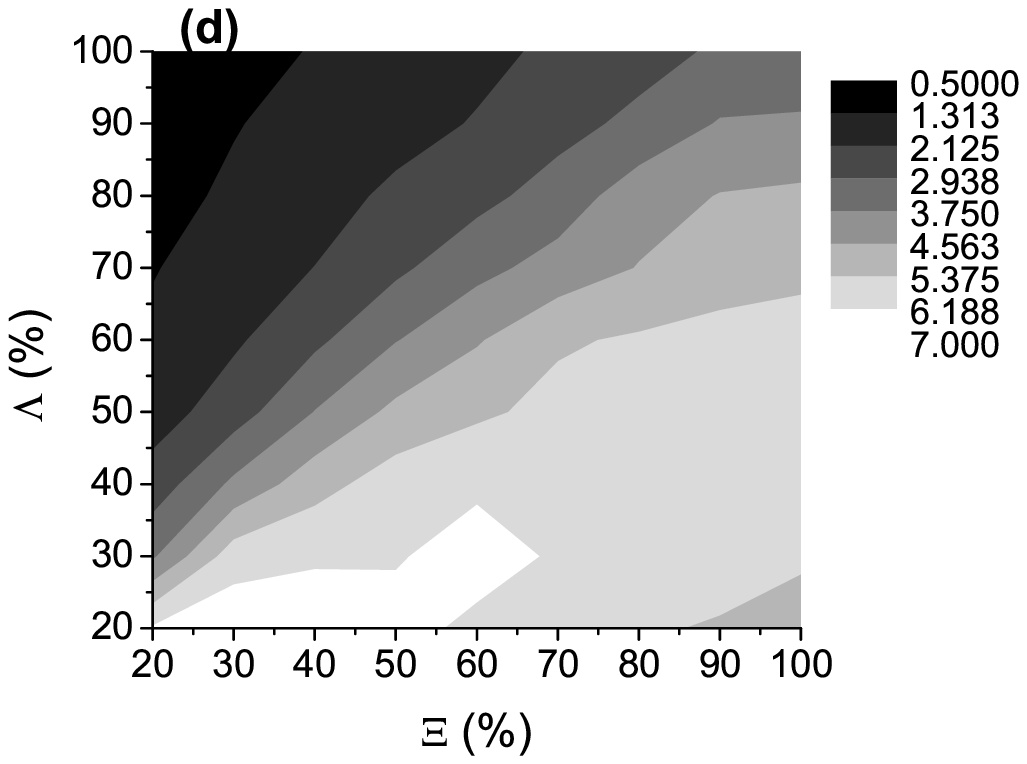}
\caption{\label{G_esc} Decay width $\Gamma$ of the ISGMR as a function of the parameters $\Xi$ and $\Lambda$ for the nuclei   (a) $^{208}\rm{Pb}$, (b) $^{144}\rm{Sm}$ ,(c) $^{116}\rm{Sn}$ and (d) $^{90}\rm{Zr}$. In (a), (b) and (c) the lines on which our results coincide with the experimental ones are also shown (solid line). The line on which the value of the decay width is maximum is shown as a dashed line.}
\end{center}
\end{figure}

\begin{figure}
\begin{center}
\includegraphics[width=9cm]{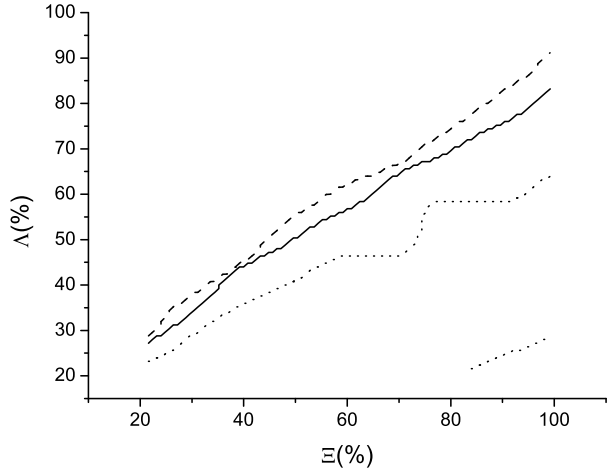}
\caption{\label{lines} The lines on the $\Xi$-$\Lambda$ plane on which our results for the decay width coincide with the experimental ones for the nuclei $^{208}\rm{Pb}$ (solid line, slope $0.63$), $^{144}\rm{Sm}$ (dashed line, slope $0.75$) and $^{116}\rm{Sn}$ (dotted lines, slopes $0.51$ and $0.46$).}
\end{center}
\end{figure}

\begin{figure}
\begin{center}
\includegraphics[width=7.5cm]{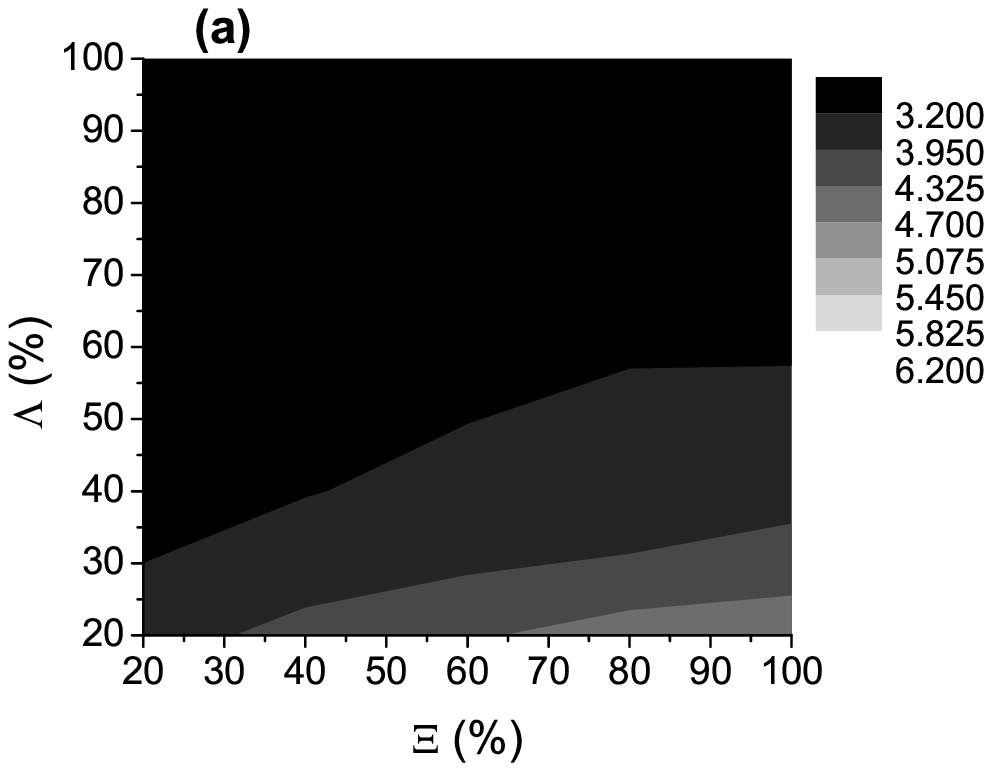}%
\includegraphics[width=7.5cm]{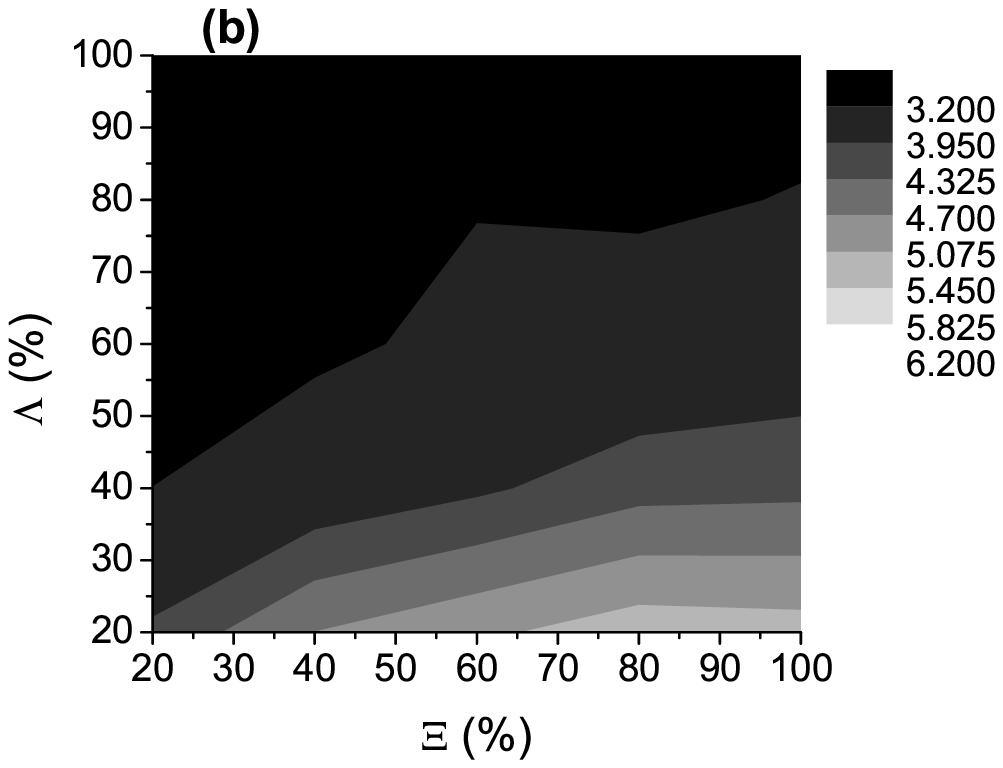}
\includegraphics[width=7.5cm]{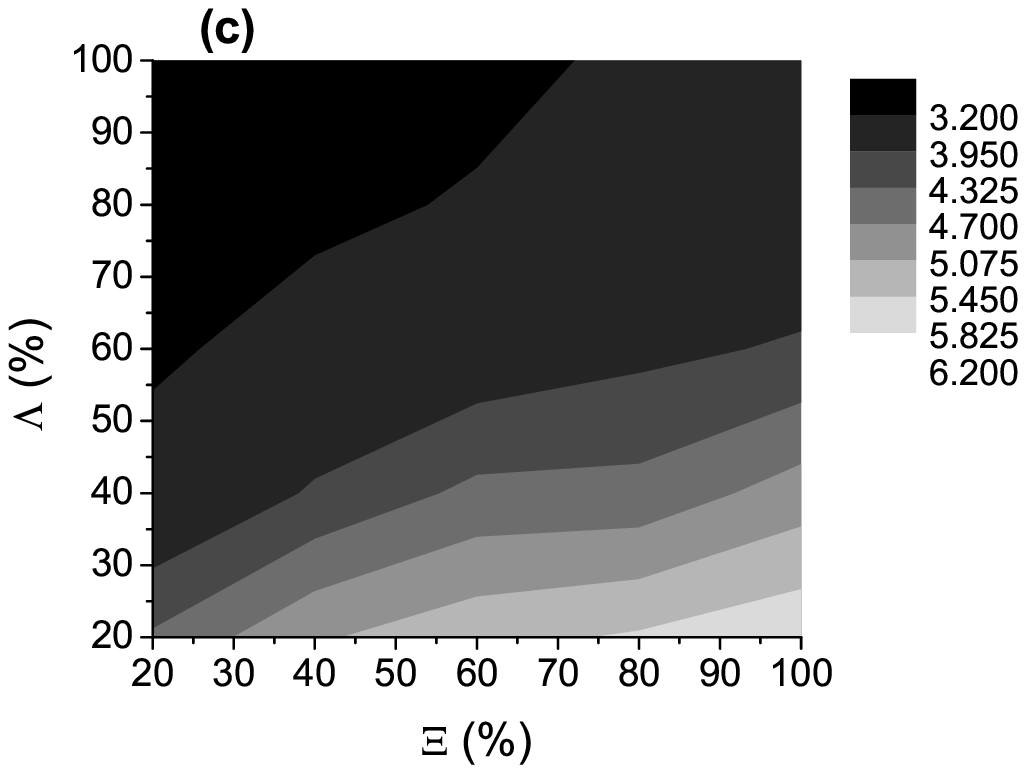}%
\includegraphics[width=7.5cm]{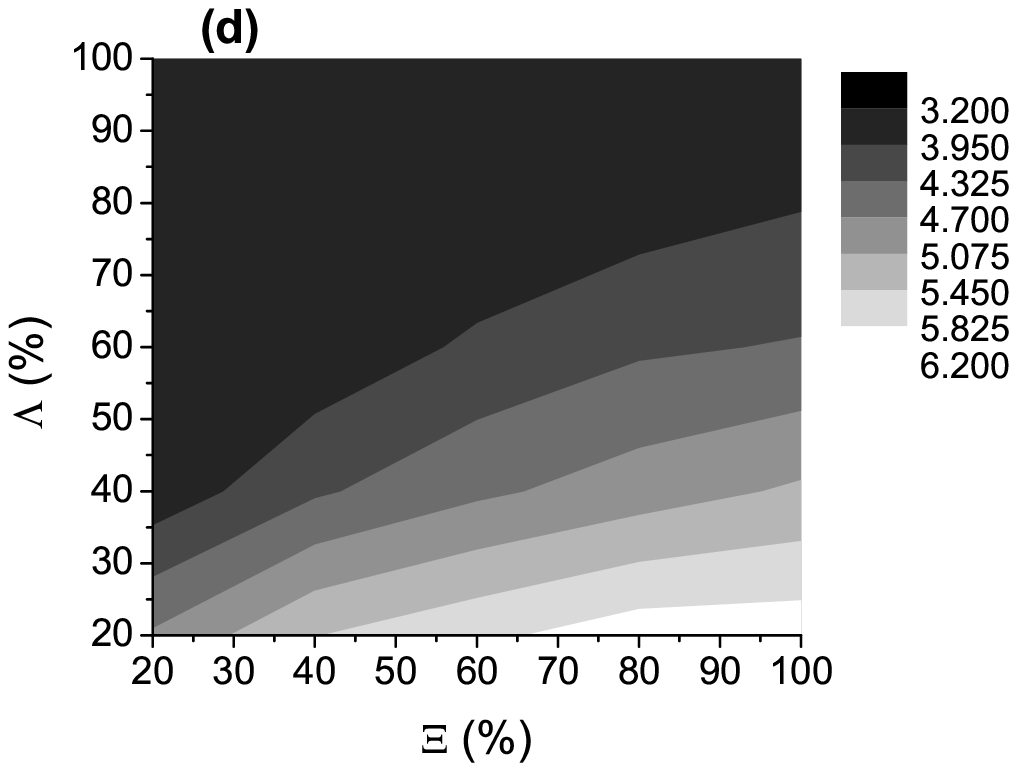}
\caption{\label{Lyap} The mean value $<\lambda_1>$ of the maximum Lyapunov exponent corresponding to the calculations of Fig.\ref{G_esc} for the nuclei (a) $^{208}\rm{Pb}$, (b) $^{144}\rm{Sm}$ ,(c) $^{116}\rm{Sn}$ and (d) $^{90}\rm{Zr}$.}
\end{center}
\end{figure}

For the nuclei $^{208}\rm{Pb}$, $^{144}\rm{Sm}$, $^{116}\rm{Sn}$ and $^{90}\rm{Zr}$, and for a $9\times 9$ grid of values for $\Xi$ and $\Lambda$, we calculate the decay width $\Gamma$ and the mean value of the largest Lyapunov exponent $<\lambda_1>$. Our results for $\Gamma$ and $<\lambda_1>$ are shown in Figs. \ref{G_esc} and \ref{Lyap} respectively. In Figs. \ref{G_esc} and \ref{lines}, the lines on the $\Xi$-$\Lambda$ plane on which our results coincide with the experimental ones are also shown for the case of the nuclei $^{208}\rm{Pb}$, $^{144}\rm{Sm}$ and $^{116}\rm{Sn}$. 

We first observe that the relationship between $<\lambda_1>$ and $\Gamma$ is in most cases what we would expect for a completely chaotic system. More specifically, for all values of $\Xi$ and $\Lambda$, both $\Gamma$ and $<\lambda_1>$ decrease as $A$ increases. Moreover, for each of the four nuclei studied, as $\Xi$ and $\Lambda$ vary, $<\lambda_1>$ and $\Gamma$ vary with the same monotonicity except from a region with $\Xi>50\%$ and $\Lambda<30\%$. Different monotonicities of $<\lambda_1>$ and $\Gamma$ have been reported in \cite{PAP05} and can be attributed to the existence of regular regions in the phase space. In all cases, the decay width $\Gamma$ seems to have its maximum value on a line with slope $\approx 1/2$ on the $\Xi-\Lambda$ plane (see Fig.\ref{G_esc}, dashed lines).

Comparing our results for the decay width with the corresponding experimental values, which are shown in Table \ref{tab_gamma}, we observe that the trend of the experimental results is reproduced, i.e. the amplitude $\Gamma$ decreases with increasing $A$ for all values of $\Xi$ and $\Lambda$. At a quantitative level, in the case of the nuclei $^{208}\rm{Pb}$ and $^{144}\rm{Sm}$, our results coincide with the experimental ones on a curve which is close to a straight line (see Figs.\ref{G_esc}(a),(b) and Fig.\ref{lines}). The corresponding slopes determined by a least squares fit equal $0.63$ and $0.75$ respectively. Above (below) these lines, our results underestimate (oveserstimate) the experimental data. In the case of the nucleus $^{116}\rm{Sn}$, our results coincide with the corresponding experimental ones on two lines with slopes $0.51$ and $0.46$ (see Fig.\ref{G_esc}(c) and Fig.\ref{lines}). For $(\Xi,\Lambda)$ points between these lines our results overestimate the experimental ones, whereas for points outside this region the experimental data are underestimated. In the case of the nucleus $^{90}\rm{Zr}$, our results for $\Gamma$ are smaller than the corresponding experimental ones for all values of $\Xi$ and $\Lambda$. It seems that our model is more successful in the description of the decay width of medium-heavy nuclei. This may be due to the decreased collectivity of the ISGMR in lighter nuclei ($A<100$) as well as to the increasing importance of higher-body processes in these nuclei. In the considered model, as it has already been mentioned, although the nucleons do not interact with each other, their motions are indirectly coupled by means of their interaction with the moving nuclear surface and therefore part and not the whole effect of the higher body processes is taken into account.

We must mention that we have also considered the case of 1 particle moving inside the potential well, as in Ref. \cite{PAP05}.
In that case, although the trend of the experimental data is reproduced ($\Gamma$ decreases with increasing $A$), the results for $\Gamma$ are very small compared to the experimental ones. Therefore, the consideration of many particles turned out necessary for the quantitative description of the decay widths.

\begin{table}
\begin{center}
\begin{tabular}{|r||c|c|c|}\hline
Nucleus & $E_x(\rm{MeV})$ & $\Gamma_{exp}(\rm{MeV})$ & $\mu_{esc}$\\
\hline
$^{208}\rm{Pb}$ & $13.96\pm 0.20$ \cite{YOU04a} & $2.88\pm 0.20$ \cite{YOU04a} & $0.11\pm 0.042$ \cite{BRA89} \\
$^{144}\rm{Sm}$ & $15.40\pm 0.30$ \cite{YOU04a}  & $3.40\pm 0.20$ \cite{YOU04a} & - \\
$^{116}\rm{Sn}$ & $15.85\pm 0.20$ \cite{YOU04a} & $5.27\pm 0.25$ \cite{YOU04a}& -\\
$^{90}\rm{Zr}$ & $17.81 +0.20 , -0.32$ \cite{YOU04} & $7.86 +0.89 ,-1.41$ \cite{YOU04} & $\simeq 0.08$ \cite{GOR00}\\
$^{112}\rm{Sn}$ & $15.67\pm 0.11$ \cite{LUI05} & $5.18 +0.40,-0.04$ \cite{LUI05} & -\\
$^{124}\rm{Sn}$ & $15.34\pm 0.13$ \cite{LUI05} & $5.00 +0.53,-0.03$ \cite{LUI05} & -\\
\hline
\end{tabular}
\caption{\label{tab_gamma} Experimental results for the excitation energy $E_x$, the decay width $\Gamma_{exp}$ and the fraction $\mu_{esc}$ of the width due to nucleon escape for the ISGMR in the nuclei $^{208}\rm{Pb}$, $^{144}\rm{Sm}$, $^{116}\rm{Sn}$, $^{90}\rm{Zr}$, $^{112}\rm{Sn}$ and $^{124}\rm{Sn}$ (Results for $^{116}\rm{Sn}$ coincide with those reported recently \cite{GAR07}).}
\end{center}
\end{table}

\begin{figure}
\begin{center}
\includegraphics[width=9cm]{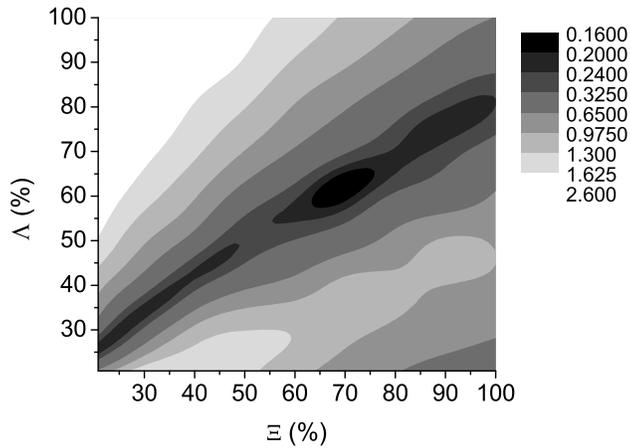}
\caption{\label{xisquare} The sum $\xi^2$ of the squares of the relative differences between our results and the experimental data for the nuclei $^{208}\rm{Pb}$, $^{144}\rm{Sm}$, $^{116}\rm{Sn}$ and $^{90}\rm{Zr}$ (see Eq.\ref{nuc_eq_ksi}).}
\end{center}
\end{figure}

\begin{figure}
\begin{center}
\includegraphics[width=9cm]{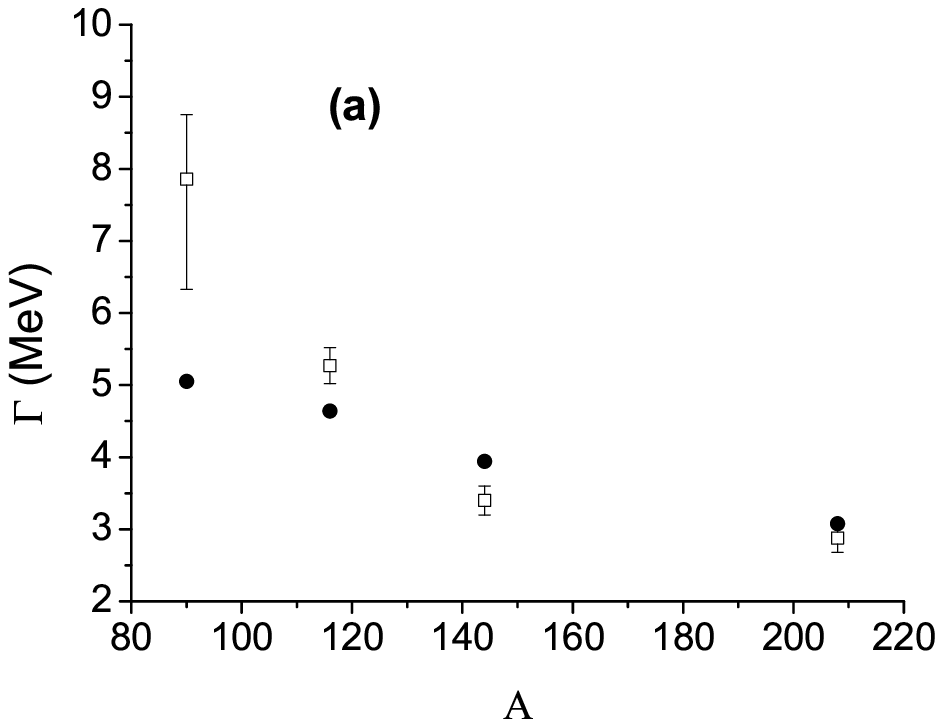}%
\includegraphics[width=9cm]{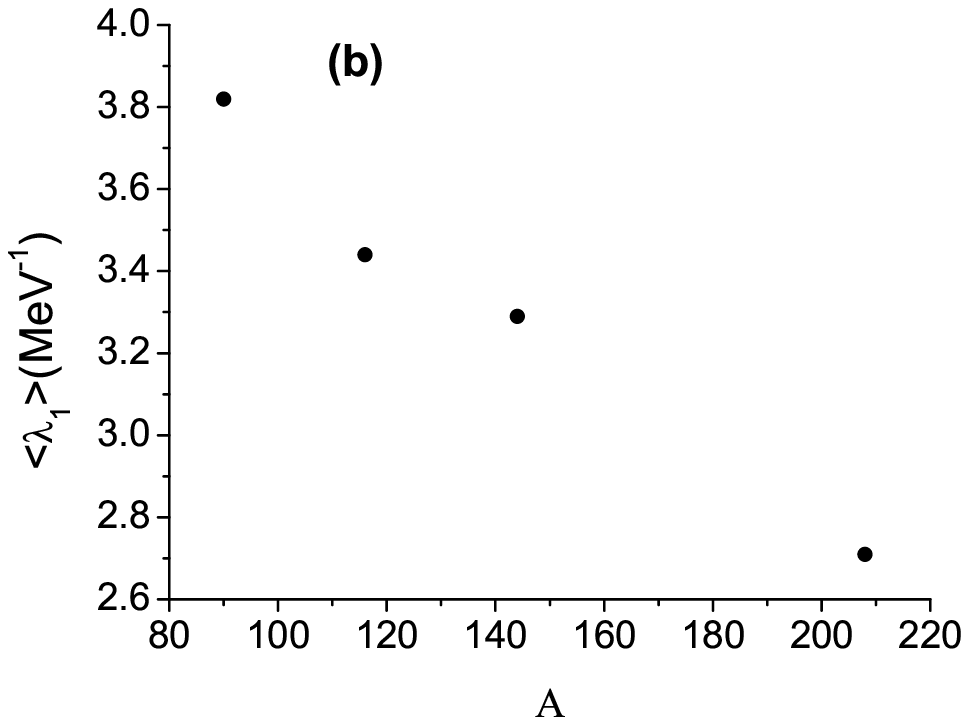}%
\caption{\label{compare} (a) The calculated values of the total decay width $\Gamma$ for the nuclei $^{208}\rm{Pb}$, $^{144}\rm{Sm}$, $^{116}\rm{Sn}$ and $^{90}\rm{Zr}$ (circles) for the values of $\Xi$ and $\Lambda$ ($\Xi\simeq 69\%$, $\Lambda\simeq 62\%$) that minimize the quantity $\xi^2$ (see Eq.\ref{nuc_eq_ksi}). The corresponding experimental results (squares) with their errors are also shown \cite{YOU04,YOU04a}. (b) The corresponding values of the mean maximum Lyapunov exponent $<\lambda_1>$.}
\end{center}
\end{figure}

In order to determine a region in the $(\Xi,\Lambda)$ plane where there is an optimal agreement between our results and the corresponding experimental values, we assume that all four nuclei can be described using the same values of $\Lambda$ and $\Xi$, i.e. all the nuclei considered exhibit the same degree of collectivity and the same fraction of nucleons that take part in the damping. We calculate the sum $\xi^2$ of the squares of the relative differences between our results and the experimental data
\begin{equation}\label{nuc_eq_ksi}
\xi^2  = \left( {\frac{{\Gamma _{Pb}  - \Gamma _{Pb(\exp )} }}{{\Gamma _{Pb(\exp )} }}} \right)^2  + \left( {\frac{{\Gamma _{Sm}  - \Gamma _{Sm(\exp )} }}{{\Gamma _{Sm(\exp )} }}} \right)^2  + \left( {\frac{{\Gamma _{Sn}  - \Gamma _{Sn(\exp )} }}{{\Gamma _{Sn(\exp )} }}} \right)^2  + \left( {\frac{{\Gamma _{Zr}  - \Gamma _{Zr(\exp )} }}{{\Gamma _{Zr(\exp )} }}} \right)^2 ,
\end{equation} 
as a function of the parameters $\Xi$ and $\Lambda$ ($\rm{Sn}\equiv ^{116}\rm{Sn}$). The results are shown in Fig. \ref{xisquare}. From this figure we observe that we have a good agreement with the experimental results in a region on the $(\Xi,\Lambda)$ plane which is around a line with a slope approximately equal to $0.69$. The optimal agreement with the experimental results is obtained where $\xi^2$ has its minimum value, i.e. for $\Xi\simeq 69\%$ and $\Lambda\simeq 62\%$. We should note that by assuming a uniform sphere oscillating with linear radial displacement and velocity field, an oscillating mass $M$ equal to $3/5$ ($0.6$) of the total mass of the nucleus is deduced (scaling model)\cite{STR82,BER94}. For the above mentioned values of $\Xi$ and $\Lambda$, the calculated decay widths for the nuclei $^{208}\rm{Pb}$, $^{144}\rm{Sm}$, $^{116}\rm{Sn}$ and $^{90}\rm{Zr}$ equal $3.08\rm{MeV}$, $3.94\rm{MeV}$, $4.64\rm{MeV}$ and $5.05\rm{MeV}$ respectively. The above results along with the corresponding experimental ones are shown in Fig.\ref{compare}(a). We observe that the discrepancy between our results and the experimental data increases as $A$ decreases. This is possibly due to our assumption that all nuclei exhibit the same degree of collectivity, which should be modified when considering lighter nuclei. We should also mention the role of the n-body processes and the gradual change of the shape of the ISGMR response which is observed for $A\leq 100$ losing its symmetric single peaked Gaussian-like character. As for the dependence of $\Gamma$ on $A$ in our model, it is dictated by the classical mechanics of the system since the mean maximum Lyapunov exponent $<\lambda_1>$ of the system has the same dependence on $A$, as can be seen from Fig.\ref{compare}(b).  Comparing our results for the width $\Gamma$ of ISGMR with those from other approaches mentioned in the Introduction, we realize that they are generally smaller (although to a different extent) than ours. For example, recent results obtained with self-consistent Hartree-Fock plus RPA approach with $\rm{SkM}^*$ effective interaction and fully taking into account the continuum \cite{PEG05} yield the following results for the width: $\Gamma(^{208}\rm{Pb})=2.5\rm{MeV}$ and $\Gamma(^{90}\rm{Zr})=3.0\rm{MeV}$ \cite{PEGPR}, whereas the values of the width $\Gamma$ obtained with the relativistic quasiparticle RPA along the lines of Ref. \cite{PAA03} using the effective interaction DDME2 derived from an effective Lagrangian with density-dependent meson-nucleon vertex function are $\Gamma(^{208}\rm{Pb})=1.54\rm{MeV}$ and $\Gamma(^{90}\rm{Zr})=2.18\rm{MeV}$ \cite{PAA06}. We realize that both sets of values of the width are in qualitative agreement with the experimental ones but they are smaller from the latter and from the values derived with our classical model. This is expected since only 1p-1h space has been considered in RPA. Recently, as mentioned in the introduction, the influence of phonon coupling terms to the damping of Giant Resonances has been studied and led to a noticeable fragmentation and increase of the width \cite{LIT07} (for instance for $^{208}\rm{Pb}$, $\Gamma$ equals 2 MeV and 3 MeV with Relativistic RPA and Relativistic RPA plus Particle-Vibration Coupling respectively).

For the above values of $\Xi$ and $\Lambda$ ($\Xi=69\%$ and $\Lambda=62\%$), we have also calculated the decay widths of the ISGMR of two other $\rm{Sn}$ isotopes, namely  $^{112}\rm{Sn}$ and $^{124}\rm{Sn}$. Our calculation yields for the widths 4.68 MeV and 4.05 MeV respectively, whereas the experimentally obtained values equal 5.18 MeV and 5.00 MeV respectively \cite{LUI05} (Recently new results for the centroid energy $E_x$ and width have been reported \cite{GAR07}. The discrepancy with the values of $E_x$ and $\Gamma$ for $^{112}\rm{Sn}$ in Table \ref{tab_gamma} has been restored; see also conclusions). In addition, for the same values of $\Xi$ and $\Lambda$, we made a prediction for the decay width of the exotic nucleus $^{132}\rm{Sn}$ for which there is increased experimental interest. Our calculation yields $\Gamma$ equal to 3.87 MeV. In our calculation, we have used as centroid energy $E_x$ the value $15.29\rm{MeV}$, which is calculated theoretically in Ref.~\cite{PER05} by performing Hartree-Fock plus Random Phase Approximation calculation using the D1S parametrization of the Gogny two-body effective interaction. We should mention that results for the decay width $\Gamma$ of the isotopes $^{112}\rm{Sn}$, $^{124}\rm{Sn}$ and $^{132}\rm{Sn}$ have been derived from other theoretical approaches (see i.e. \cite{PAA03,LIT07,PER05,FAR97,PEG04}). The values for the decay width of $^{132}\rm{Sn}$ derived by means of the two approaches reported above equal $3.21\rm{MeV}$ and $2.13\rm{MeV}$ respectively \cite{PEGPR,PAA06}. Recently, a decay width of $^{132}\rm{Sn}$ equal to $3.09\rm{MeV}$ has been reported in \cite{LIT07} by fitting the theoretical strength distribution with a Lorentzian. These results are smaller than the value found in our calculations.

\begin{figure}
\begin{center}
\includegraphics[width=7.5cm]{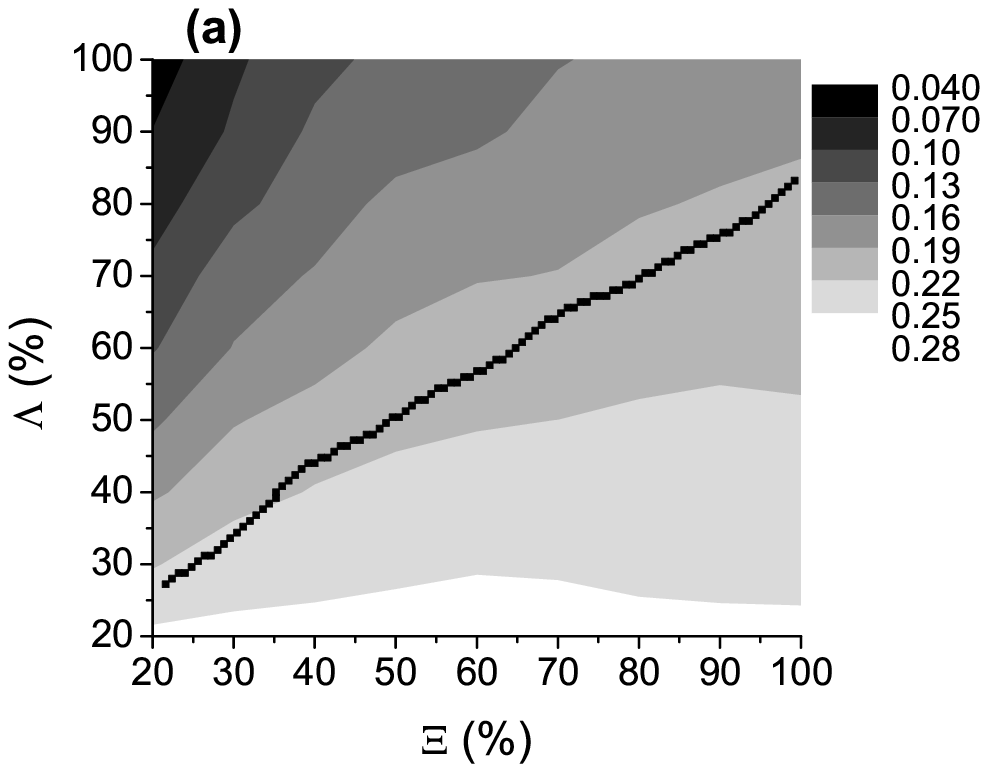}%
\includegraphics[width=7.5cm]{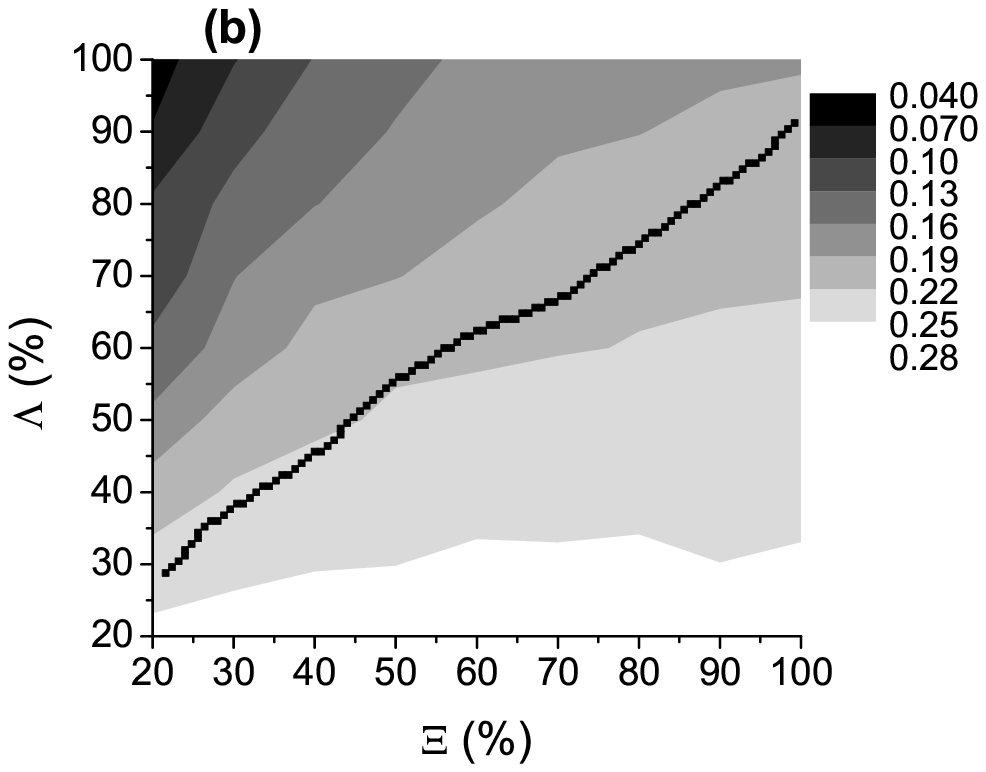}
\includegraphics[width=7.5cm]{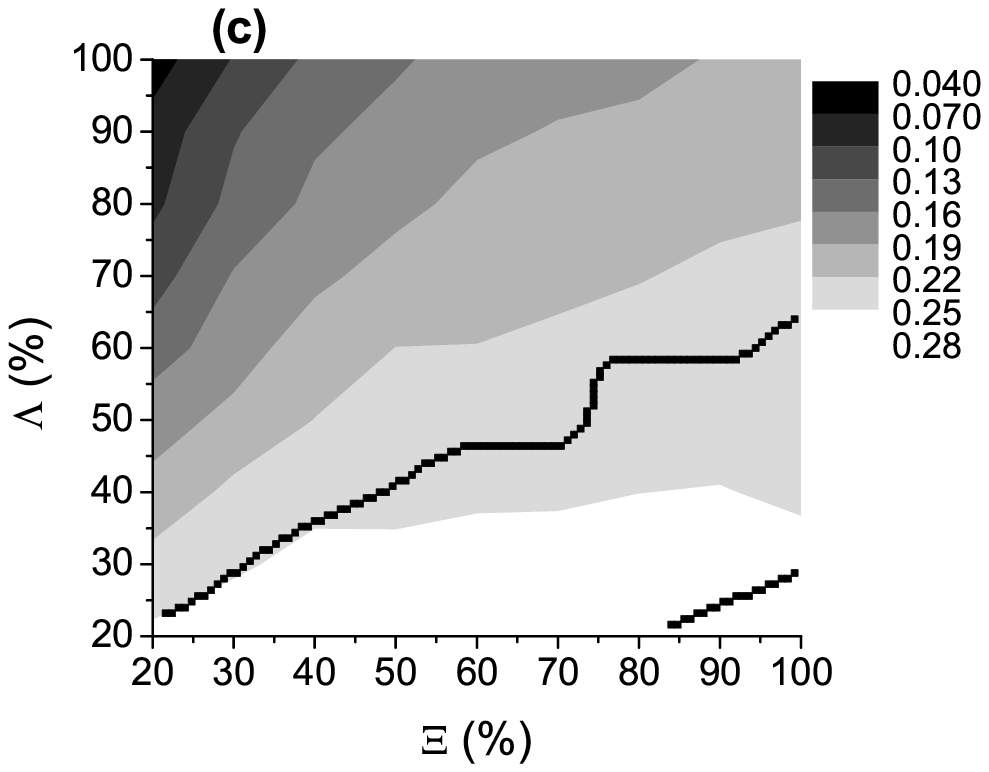}%
\includegraphics[width=7.5cm]{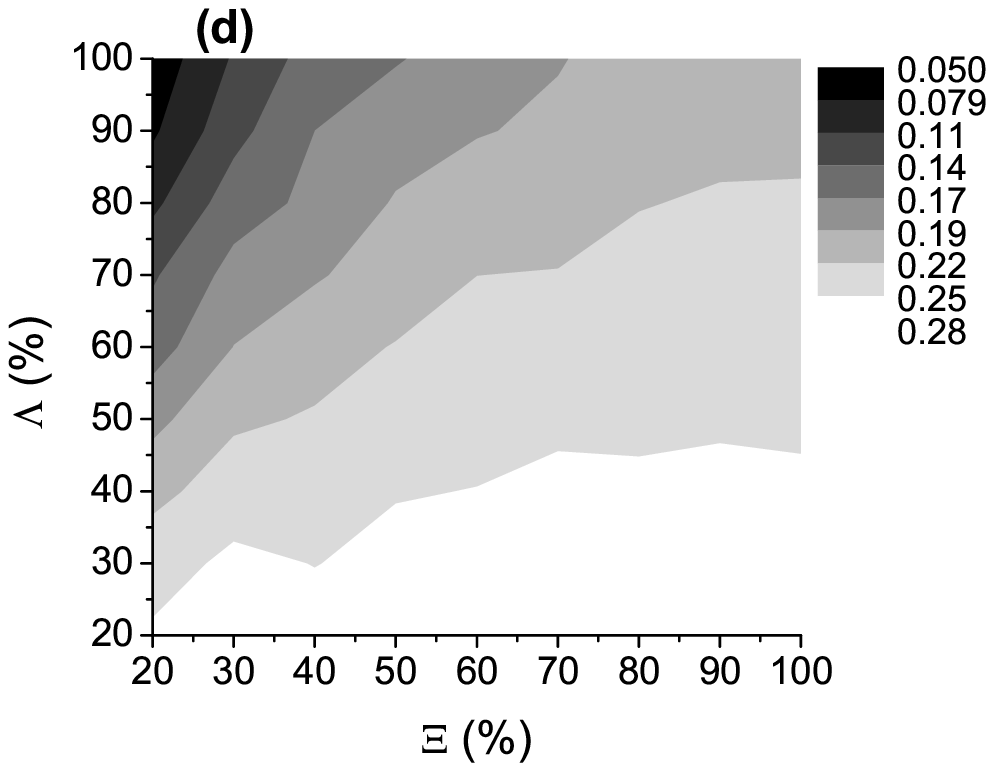}
\caption{\label{muesc} The fraction $\mu_{esc}$ of the total decay width due to the escape of nucleons as a function of the parameters $\Xi$ and $\Lambda$ for the nuclei (a) $^{208}\rm{Pb}$, (b) $^{144}\rm{Sm}$ ,(c) $^{116}\rm{Sn}$ και (d) $^{90}\rm{Zr}$.}
\end{center}
\end{figure}

We have also calculated the fraction $\mu_{esc}$ of the total decay width which is due to the escape of nucleons. It is defined as
\begin{equation}
\mu _{esc}  = \frac{{\Gamma  - \Gamma '}}{\Gamma },
\end{equation}
where $\Gamma$ is the total decay width and $\Gamma '$ is the decay width without taking into account the escape of particles. The latter is extracted from the mean value $<R(t)>$ of the collective variable over the ensemble of orbits that do not lead to nucleon escape, i.e. $\left\langle {R(t)} \right\rangle  = \sum\limits_j {R_j (t)/{\rm N}_{{\rm tot}} }$ where $R_j (t)$ is the time series of the collective variable in the $j-$th orbit and the summation is performed only over the orbits in which all the nucleons remain confined in the potential well at time $t$. In the orbits where a nucleon escapes from the potential well, the motion of the escaping nucleon is followed until its distance from the nucleus reaches a cutoff value, which has been taken much larger than the radius of the nucleus. Then, the two degrees of freedom describing the motion of the escaping nucleon are considered to be decoupled from the motion of the nucleons confined in the potential well, i.e. the escaping nucleon moves away from the nucleus with constant energy and it is therefore excluded from the sum of eq 5. Once the nucleon escapes and travels at constant energy away from the nucleus, it carries energy away from the system, thus contributing to the dissipation of the collective degree of freedom. The results are shown in Fig.\ref{muesc}. From this Figure, it can be seen that in the region of $\Xi$ and $\Lambda$ values in which agreement of our results for $\Gamma$ with the experimental data is optimal, $\mu_{esc}$ is between $0.20$ and $0.28$ for all nuclei under consideration. These values are larger than the corresponding experimental ones which are available only for the nuclei $^{208}\rm{Pb}$ and $^{90}\rm{Zr}$ (see Table \ref{tab_gamma};\cite{BRA88,BOR98,BRA89,GOR00,BOR89}. Inclusion of a Coulomb barrier leads to a decrease of $\mu_{esc}$ by a fraction ranging from $\approx 21\%$ (for $^{90}\rm{Zr}$) to $\approx 35\%$ (for $^{208}\rm{Pb}$). However, as our calculation is purely classical it does not account for the quantum tunneling of particles through the potential barrier. One should add that the experimental values for $\mu_{esc}$ reported above refer to the direct escape width whereas our model probably includes part of the statistical decay width. The partial widths for direct neutron escape of the ISGMR of $^{90}\rm{Zr}$, $^{124}\rm{Sn}$ and $^{208}\rm{Pb}$ have also been calculated recently \cite{GOR00} within a continuum extended  RPA approach with the use of a phenomenological mean field, the Landau-Migdal p-h interaction and some partial self-consistent conditions. 
\newpage
\section{Conclusions and prospects}
In conclusion, in this work we studied the decay of the Isoscalar Giant Monopole Resonance (ISGMR) for several spherical nuclei using a classical model, in which a number of noninteracting nucleons (particles) move in an oscillating potential well and can exchange energy with the nuclear surface (wall of the potential well). The motion of the nuclear surface is described by a collective variable which appears explicitly in the Hamiltonian of the system as an additional degree of freedom. The total energy of the system is therefore conserved. Since there is no interaction between the particles, the model takes explicitly into account only one-body processes. Nevertheless, the particles interact indirectly with each other by means of the coupling of their motion to the motion of the oscillating well and therefore dissipation beyond the one-body is expected to be taken into account. Moreover, for medium-heavy nuclei, with the proper choice of the free parameters of the system (the degree of collectivity and the fraction of nucleons that participate to the decay), our results show a good agreement with the experimental data, at least comparable to the agreement of more sophisticated quantum models. The dependence of the decay width of the ISGMR on the size of the nucleus has been found to be related to the characteristics of the underlying classical dynamics, i.e. to the maximum Lyapunov exponent. Its value is determined by the presence of chaotic regions in the phase space. Chaoticity is included in the dynamics of GRs due to the time-dependence and to the residual interaction. We expect that at least qualitatively, our results should hold in the quantum systems. Recently, there have been experimental results for the ISGMR of several isotopes of Sn from RCNP in Osaka \cite{GAR07}. The theoretical calculations of the energy give values that are larger than the experimental ones. These imply larger values of the compression moduli. Work is in progress to investigate with our model in more detail the underlying classical dynamics of ISGMR in these isotopes that differ in the symmetry energy and predict their decay widths. We should add that our model properly adjusted can be used for the study of the decay of ISGMR of other nuclei including those far from stability as well as for the study of the Isovector Giant Monopole Resonances.

\begin{acknowledgments}
The authors acknowledge financial support through the research program "Pythagoras" of the EPEAEK II of the European Union and the Greek Ministry of Education  as well as through the grant 70/4/3309 of the University of Athens. Support through DGF Sonderforschungsbereich SFB 634 is also acknowledged. P.P. would also like to thank the Greek Scholarships Foundation (IKY) for financial support. Useful correspondence with M. Baldo is acknowledged.
\end{acknowledgments}

\end{document}